\documentclass[superscriptaddress,aps,prd,nofootinbib,twocolumn,showpacs]{revtex4-1}
\usepackage{amsfonts}
\usepackage{amsmath}
\usepackage{amssymb}
\usepackage{graphicx}
\usepackage{mathrsfs}
\usepackage{hhline,color}
\usepackage{pifont,ulem}
\usepackage{lmodern} 
\usepackage[utf8]{inputenc}

\newcommand{\unit}[1]{\ensuremath{\, \mathrm{#1}}}
\newcommand{\pr}[1]{\ensuremath{\left[#1\right]}}
\newcommand{\pc}[1]{\ensuremath{\left(#1\right)}}
\newcommand{\chav}[1]{\ensuremath{\left\{#1\right\}}}
\newcommand{\ev}[1]{\ensuremath{\left\langle #1\right\rangle}}

\def\beq{\begin{equation}}
\def\eeq{\end{equation}}
\def\beqa{\begin{eqnarray}}
\def\eeqa{\end{eqnarray}}
\def\ban{\begin{eqnarray*}}
\def\ean{\end{eqnarray*}}
\def\bi{\begin{itemize}}
\def\ei{\end{itemize}}

\newcommand{\Z}{\mathbb{Z}}

\begin{document}

\title{Inverse magnetic catalysis in the (2+1)-flavor Nambu--Jona-Lasinio and Polyakov--Nambu--Jona-Lasinio models }

\author{M. Ferreira}
\affiliation{Centro de F\'{\i}sica Computacional, Department of Physics,
University of Coimbra, P-3004 - 516  Coimbra, Portugal}
\affiliation{Instituto Tecnol\'ogico de Aeron\'autica, 12.228-900, S\~ao Jos\'e dos Campos, SP, Brazil}
\author{P. Costa}
\affiliation{Centro de F\'{\i}sica Computacional, Department of Physics,
University of Coimbra, P-3004 - 516  Coimbra, Portugal}
\author{O. Louren\c co}
\affiliation{Departamento de Ci\^encias da Natureza, Matem\'atica e Educa\c
c\~ao, CCA, Universidade Federal de S\~ao Carlos, 13600-970, Araras, SP, Brazil}
\author{T. Frederico}
\affiliation{Instituto Tecnol\'ogico de Aeron\'autica, 12.228-900, S\~ao Jos\'e dos Campos, SP, Brazil}
\author{C. Provid\^encia}
\affiliation{Centro de F\'{\i}sica Computacional, Department of Physics,
University of Coimbra, P-3004 - 516  Coimbra, Portugal}

\date{\today}

\begin{abstract}
The QCD phase diagram at zero chemical potential and finite
temperature subject to an external magnetic field is studied within the 
three-flavor Nambu--Jona-Lasinio (NJL) model  and the NJL model with the
Polyakov loop. A scalar coupling parameter dependent on the
magnetic field intensity is considered. The scalar coupling has been
fitted so that the lattice QCD pseudocritical chiral transition
temperatures are reproduced and in the limit of large magnetic field
decreases with the inverse of the magnetic field intensity. This
dependence of the coupling allows us to reproduce the lattice QCD results
with respect to the quark condensates and Polyakov loop:  due to the
magnetic field the quark
condensates  are enhanced at low and high
temperatures and suppressed for
temperatures close to the transition temperatures and the Polyakov loop
increases with the magnetic field.
\end{abstract}


\maketitle

\section{Introduction}

In the last years, magnetized quark matter has attracted the attention of the physics community due
to its  relevance for heavy ion collisions at very high energies \cite{HIC,kharzeev},
to the understanding of the first phases of the Universe \cite{cosmo} and for studies involving compact
objects like magnetars \cite{duncan}.

In the presence of an external magnetic field, the behavior of quark matter is determined by  the
competition between two different mechanisms: the enhancement of the quark condensate because of
the opening of the gap between the Landau states leading to the increase of low-energy contributions 
to the formation of the chiral condensate; and the suppression of the quark condensate due to the 
partial restoration of chiral symmetry.
It was shown that in the region of  low momenta relevant for chiral symmetry breaking there is a strong
screening effect of the gluon interactions,  which suppresses the condensate 
\cite{Miransky:2002rp,endrodi2013}. In this region, the gluons acquire a mass $M_g$
of the order of $\sqrt{N_f \alpha_s|eB|}$, due to the coupling of the gluon field to a
quark-antiquark  interacting state. 
In the presence of a strong enough magnetic field, this mass $M_g$ for gluons
becomes larger.
This, along with the property that the strong coupling $\alpha_s$ decreases with increasing $eB$, 
$\alpha_s(eB)=(b\ln(|eB|/\Lambda_{QCD}^2))^{-1}$ with $b=(11N_c-2N_f)/6\pi=27/6\pi$, \cite{Miransky:2002rp}
leads  to an effective weakening of the  interaction between the quarks in the presence of an external
magnetic field, and damps  the chiral condensate. 

The suppression of the quark condensate, also known as inverse magnetic catalysis (IMC), 
manifests itself on the decrease of the pseudocritical chiral transition temperature 
obtained in lattice QCD (LQCD) calculations with physical quark masses and an
increase of the Polyakov field \cite{baliJHEP2012,bali2012PRD,endrodi2013}. 
Recent results in two-flavor LQCD with dynamical overlap fermions 
in an external magnetic field also support the IMC scenario \cite{Bornyakov:2013eya}.
In particular,  in \cite{endrodi2013} it is argued that the IMC 
may be a consequence  of how the gluonic sector reacts to the presence of a magnetic
field, and,   it is shown that the magnetic field drives up the expectation value of 
the Polyakov field. The distribution of gluon fields changes as a consequence of the 
distortion of the quark loops in the magnetic field background. 
Therefore, the backreaction of the quarks  on the gauge fields should be incorporated in  
effective models in order to describe the IMC.

Almost all low-energy effective models, at zero chemical potential, including the 
Nambu--Jona-Lasinio (NJL)-type models, find an enhancement of the condensate due 
to the magnetic field, and no reduction of the pseudocritical chiral transition
temperature with the magnetic field \cite{reviews}.
However, a recent study using the Polyakov--Nambu--Jona-Lasinio (PNJL) \cite{Ferreira:2013tba} 
has shown that if the LQCD data \cite{bali2012PRD} is fitted by making the pure-gauge critical
temperature $T_0$, a parameter of the PNJL model, $eB$ dependent, the model is able to 
describe an IMC.
Several recent  studies 
\cite{Fukushima:2012kc,Chao:2013qpa,Andersen:2013swa,Kamikado:2013pya,Farias:2014eca,Fraga:2013ova} 
discuss the origin of the IMC phenomenon. 
In particular, the magnetic inhibition can be a possible explanation for the decreasing behavior of
the chiral restoration temperature with increasing $eB$ \cite{Fukushima:2012kc};
another mechanism to explain the IMC around the 
critical temperature as induced by sphalerons was proposed in \cite{Chao:2013qpa}.

The discussion above points out that the gluon distribution reacts to the magnetic background
and suggests that  the effective interaction between the quarks   should get this dependence. 
With this motivation,  we adopt a running coupling constant of the chiral invariant quartic 
quark interaction in NJL and PNJL models with the magnetic field.  The damping of the strength 
of the effective quartic interaction is built phenomenologically, keeping SU(3) flavor symmetry,  
under different assumptions  inspired by lattice results for the quark condensate at finite 
temperature and magnetic field.
 
This paper is organized as follows. In Sec. II, we briefly present the PNJL model used in 
this work, the Polyakov loop potential, and the parametrizations chosen. In Sec. III, 
the importance of the running coupling in the (P)NJL models for magnetized quark matter is discussed. 
Also, the behavior of the condensates with temperature and the magnetic field
intensity is compared  with the LQCD results. Finally, in Sec. IV, the main conclusions are drawn.


\section{Model and Formalism}
\label{sec:model}

The PNJL Lagrangian with explicit chiral symmetry breaking, where the quarks couple 
to a (spatially constant) temporal background gauge field, represented in terms of the
Polyakov loop, and in the presence of an external magnetic field is given by 
\cite{PNJL}
\begin{eqnarray}
{\cal L} &=& {\bar{q}} \left[i\gamma_\mu D^{\mu}-
	{\hat m}_c \right ] q ~+~ {\cal L}_{sym}~+~{\cal L}_{det} \nonumber\\
&+& \mathcal{U}\left(\Phi,\bar\Phi;T\right) - \frac{1}{4}F_{\mu \nu}F^{\mu \nu},
	\label{Pnjl}
\end{eqnarray}
where the quark sector is described by the  SU(3) version of the NJL model which
includes scalar-pseudoscalar and the 't Hooft six fermion interactions that
models the axial $U_A(1)$ symmetry breaking  \cite{Hatsuda:1994pi.Klevansky:1992qe},
with ${\cal L}_{sym}$ and ${\cal L}_{det}$  given by \cite{Buballa:2003qv},
\begin{eqnarray}
	{\cal L}_{sym}= \frac{G_s}{2} \sum_{a=0}^8 \left [({\bar q} \lambda_ a q)^2 + 
	({\bar q} i\gamma_5 \lambda_a q)^2 \right ] ,
\end{eqnarray}
\begin{eqnarray}
	{\cal L}_{det}=-K\left\{{\rm det} \left [{\bar q}(1+\gamma_5)q \right] + 
	{\rm det}\left [{\bar q}(1-\gamma_5)q\right] \right \}
\end{eqnarray}
where $q = (u,d,s)^T$ represents a quark field with three flavors, 
${\hat m}_c= {\rm diag}_f (m_u,m_d,m_s)$ is the corresponding (current) mass matrix,
$\lambda_0=\sqrt{2/3}I$  where $I$ is the unit matrix in the three-flavor space, 
and $0<\lambda_a\le 8$ denote the Gell-Mann matrices.
The coupling between the (electro)magnetic field $B$ and quarks, and between the 
effective gluon field and quarks is implemented  {\it via} the covariant derivative 
$D^{\mu}=\partial^\mu - i q_f A_{EM}^{\mu}-i A^\mu$ where $q_f$ represents the 
quark electric charge ($q_d = q_s = -q_u/2 = -e/3$),  $A^{EM}_\mu$ and 
$F_{\mu \nu }=\partial_{\mu }A^{EM}_{\nu }-\partial _{\nu }A^{EM}_{\mu }$ 
are used to account for the external magnetic field and 
$A^\mu(x) = g_{strong} {\cal A}^\mu_a(x)\frac{\lambda_a}{2}$ where
${\cal A}^\mu_a$ is the SU$_c(3)$ gauge field.
We consider a  static and constant magnetic field in the $z$ direction, 
$A^{EM}_\mu=\delta_{\mu 2} x_1 B$.
In the Polyakov gauge and at finite temperature the spatial components of the 
gluon field are neglected: 
$A^\mu = \delta^{\mu}_{0}A^0 = - i \delta^{\mu}_{4}A^4$. 
The trace of the Polyakov line defined by
$ \Phi = \frac 1 {N_c} {\langle\langle \mathcal{P}\exp i\int_{0}^{\beta}d\tau\,
A_4\left(\vec{x},\tau\right)\ \rangle\rangle}_\beta$
is the Polyakov loop which is the order parameter of the $\Z_3$ 
symmetric/broken phase transition in pure gauge.

To describe the pure-gauge sector an effective potential $\mathcal{U}\left(\Phi,\bar\Phi;T\right)$
is chosen in order to reproduce the results obtained in lattice calculations \cite{Roessner:2006xn},
\begin{eqnarray}
	& &\frac{\mathcal{U}\left(\Phi,\bar\Phi;T\right)}{T^4}
	= -\frac{a\left(T\right)}{2}\bar\Phi \Phi \nonumber\\
	& &
	+\, b(T)\mbox{ln}\left[1-6\bar\Phi \Phi+4(\bar\Phi^3+ \Phi^3)-3(\bar\Phi \Phi)^2\right],
	\label{Ueff}
\end{eqnarray}
where $a\left(T\right)=a_0+a_1\left(\frac{T_0}{T}\right)+a_2\left(\frac{T_0}{T}\right)^2$, 
$b(T)=b_3\left(\frac{T_0}{T}\right)^3$.
The standard choice of the parameters for the effective potential $\mathcal{U}$ is
$a_0 = 3.51$, $a_1 = -2.47$, $a_2 = 15.2$, and $b_3 = -1.75$.   
The value of $T_0=210\unit{MeV}$ is fixed in order to reproduce LQCD results 
($\sim$ 170 MeV \cite{Aoki:2009sc}),

We use as a regularization scheme, a sharp cutoff, $\Lambda$, in three-momentum space, 
only for the divergent ultraviolet sea quark integrals.  
The parameters of the model, $\Lambda$, the coupling constants $G_s$ and $K$,
and the current quark masses $m_u$ and $m_s$ are determined  by fitting
$f_\pi$, $m_\pi$ , $m_K$ and $m_{\eta'}$ to their empirical values. 
We consider $\Lambda = 602.3$, MeV, $m_u= m_d=5.5$, MeV,
$m_s=140.7$ MeV, $G_s \Lambda^2= 3.67$ and $K \Lambda^5=12.36$
as in \cite{Rehberg:1995kh}.
The thermodynamical potential for the three-flavor quark sector $\Omega$ is written as
\begin{align}
\Omega(T,\mu)&=G_s\sum_{f=u,d,s}\ev{\bar{q}_fq_f}^2
+4K\ev{\bar{q}_uq_u}\ev{\bar{q}_dq_d}\ev{\bar{q}_sq_s} \nonumber \\
+&{\cal U}(\Phi,\bar{\Phi},T)+\sum_{f=u,d,s}\pc{\Omega_{\text{vac}}^f+\Omega_{\text{med}}^f
+\Omega_{\text{mag}}^f}
\end{align}
where the flavor contributions from vacuum $\Omega^{\text{vac}}_f$, medium $\Omega^{\text{med}}_f$, 
and magnetic field $\Omega^{\text{mag}}_f$ \cite{Menezes:2008qt.Menezes:2009uc} are given by
\begin{align}
 \Omega_{\text{vac}}^f&=-6\int_{\Lambda}\frac{d^3p}{(2\pi)^3}\sqrt{p^2+M_f^2}\\
\Omega_{\text{med}}^f&=-T\frac{|q_fB|}{2\pi}\sum_{k=0}\alpha_k\int_{-\infty}^{+\infty}\frac{dp_z}{2\pi}\pc{Z_{\Phi}^+(E_f)+Z_{\Phi}^-(E_f)}\\
 \Omega_{\text{mag}}^f&=-\frac{3(|q_f|B)^2}{2\pi^2}\pr{\zeta^{'}(-1, x_f)-\frac{1}{2}(x_f^2-x_f)\ln x_f+\frac{x_f^2}{4}} 
\end{align}
where $E_f=\sqrt{p_z^2+M_f^2+2|q_f|Bk}$ , $\alpha_0=1$ and $\alpha_{k>0}=2$,
$x_f=M_f^2/(2|q_f|B)$, and $\zeta^{'}(-1, x_f)=d\zeta(z, x_f)/dz|_{z=-1}$, 
where $\zeta(z, x_f)$ is the Riemann-Hurwitz zeta function. 
At zero chemical potential the quark distribution functions 
$Z_{\Phi}^+(E_f)$ and $Z_{\Phi}^-(E_f)$ read
\begin{align}
 Z_{\Phi}^+=Z_{\Phi}^-=\ln\chav{1+3\Phi e^{-\beta E_f}+3\Phi e^{-2\beta E_f}+e^{-3\beta E_f}}
\label{eq:z}
\end{align}
once $\bar{\Phi}=\Phi$. 

\section{Running coupling in the (P)NJL Model for Magnetized Quark Matter}
\label{sec:Running}

\subsection{NJL model}

As already referred, the presence of an external magnetic field has two competing
mechanisms: on  one hand it enhances the  chiral condensate due to the increase 
of low-energy contributions; on the other hand there is a suppression of the condensate
because in the region of the low momenta relevant for the chiral symmetry breaking 
mechanism there is a strong screening effect of the gluon interactions 
\cite{Miransky:2002rp,endrodi2013}. 
This suppression of the condensate, also known as IMC, manifests 
itself as the  decrease of the pseudocritical chiral transition temperature obtained in 
LQCD calculations with physical quark masses \cite{baliJHEP2012,bali2012PRD} and in the 
increasing of the Polyakov loop \cite{endrodi2013}.

Within the NJL and PNJL models the inclusion of the magnetic field in
the Lagrangian density allows us to describe  the magnetic catalysis
effect, but fails to account for the IMC.  
In the NJL model the quarks interact through local current-current couplings,
assuming that the gluonic degrees of freedom can be frozen into pointlike 
effective interactions between quarks. Therefore, we may expect that the screening 
of the gluon interaction discussed above weakens the interaction and which is
translated into a decrease of the scalar coupling with the intensity of the 
magnetic field.

\begin{figure}[t]
\centering
    \includegraphics[width=0.85\linewidth,angle=0]{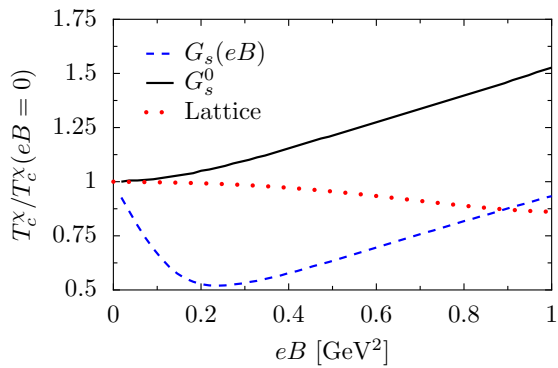}
\caption{
The renormalized critical temperatures
of the chiral transition ($T_c^\chi(eB=0)=178$ MeV) as a function of $eB$ in the NJL model with a
magnetic field dependent coupling $G_s(eB)$ (blue dashed) and a constant
coupling $G_s^0$ (black line),  
and the lattice results (red dots) \cite{baliJHEP2012}.
}
\label{fig:temp_crit}
\end{figure}

In \cite{Miransky:2002rp}, it was shown that the running coupling decreases with
the magnetic field strength,
\begin{equation}
\alpha_s(eB)=\frac{1}{b\ln\frac{|eB|}{\Lambda_{QCD}^2}}
\end{equation}
with $b=(11N_c-2N_f)/6\pi=27/6\pi$.\\
Consequently, in the NJL model the coupling $G_s$, which can be seen as $\propto\alpha_s$, 
must decrease with an increasing magnetic field strength. 

A first attempt to include the impact of the running coupling in the NJL model can be done
by introducing the simple ansatz, 
\footnote{When we were finalizing this article, the same idea was implemented in the
SU(2) version of the NJL model \cite{Farias:2014eca}. However, in this paper, besides
we are dealing with the SU(3) version of the model, we will fit $G_s(eB)$ to the
LQCD results for the chiral transition pseudocritical temperature.}
\begin{equation} 
G_s(eB)=G_s^0/\ln\pc{e+|eB|/\Lambda_{QCD}^2}.
\label{geB1}
\end{equation}
In the limit case $eB\rightarrow\infty$, we obtain $G_s\rightarrow0$, and for
$eB\rightarrow0$, we get $G_s\rightarrow G_s(eB=0)=G_s^0$. 
The pseudocritical temperatures for the chiral transitions
{$T^\chi_c=(T^\chi_u+T^\chi_d)/2$ (being $T^\chi_u$ and $T^\chi_d$
the transition temperatures for $u$ and $d$ quarks, respectively)},
are calculated using the location of the susceptibility peaks,
$C_f = -m_\pi\partial\sigma_f/\partial T$, with 
$\sigma_f=\left\langle \bar{q_f}q_f\right\rangle(B,T)/\left\langle \bar{q_f}q_f\right\rangle(0,0)$.
The multiplication by $m_\pi$ is only to ensure that the susceptibilities are 
dimensionless.
Other methods to define the temperature transitions, such as those from the
magnitude of the order parameters, are equally useful, see, for instance,
Ref. \cite{odilon}.
The calculated chiral pseudocritical temperatures are shown in Fig. \ref{fig:temp_crit}:
when $G_s=G_s^0$ the model always shows a magnetic catalyzes, with increasing $T_c^{\chi}/T_c^{\chi}(eB)$
for all range of magnetic fields;  
when $G_s=G_s(eB)$, defined by Eq. (\ref{geB1}), an IMC is seen until 
$eB \approx 0.3$ GeV$^2$, 
with the decrease of the pseudocritical temperature for low magnetic fields, but 
with the increase of $T_c^{\chi}/T_c^{\chi}(eB)$   for high $eB$  values. 
Thus, with this simple ansatz, the model predicts an IMC at low $eB$ 
and magnetic catalysis at high $eB$. 
This is in agreement with lattice results at high $eB$ \cite{Ilgenfritz:2013ara,D'Elia:2010nq}.
It is worth noting that the logarithm behavior of the running coupling of QCD 
$\alpha_s(p^2)$, occurs for high  momentum transfers $p>>$ 1 GeV.
In this way, the $\alpha_s(eB)\propto\ln\pc{|eB|/\Lambda_{QCD}^2}^{-1}$ behavior, 
may not be suitable for the low magnetic field range, $eB<1$ GeV$^2$. \\

\begin{figure}[t!]
\centering
    \includegraphics[width=0.85\linewidth,angle=0]{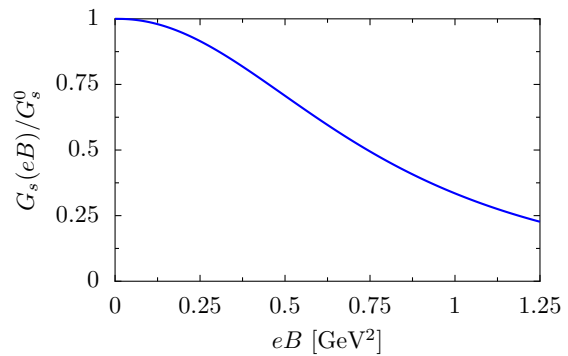}
\caption{
The $G_s(eB)$ dependence calculated in the NJL model in order
to reproduce the LQCD renormalized chiral transition temperature \cite{baliJHEP2012}.
}
\label{fig:fit}
\end{figure}

\begin{figure}[th!]
\centering
    \includegraphics[width=0.85\linewidth,angle=0]{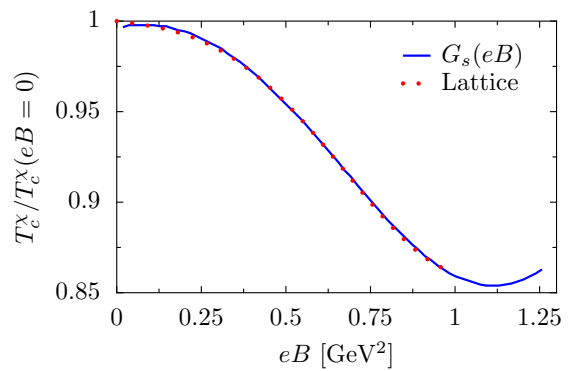}
\caption{
The renormalized critical temperature
of the chiral transition as a function of $eB$ in the NJL model, with the
magnetic field dependent coupling $G_s(eB)$ [Eq. (\ref{eq:fit})] (blue line)  
and LQCD results (red dots) \cite{baliJHEP2012}.
}
\label{Fig:T_c}
\end{figure}

Since there is no LQCD data for $\alpha_s(eB)$ available, we will use another approach, 
in particular we will fit $G_s(eB)$ in order to reproduce $T_c^\chi(eB)$ obtained in 
LQCD calculations \cite{baliJHEP2012}. 
The resulting fit function of $G_s(eB)$, that reproduces the $T_c^\chi(eB)$ 
(see Fig. \ref{Fig:T_c}), is plotted in Fig. \ref{fig:fit} and is written as
\begin{equation}
G_s(\zeta)=G_s^0\pc{\frac{1+a\,\zeta^2+b\,\zeta^3}
{1+c\,\zeta^2+d\,\zeta^4}}\,
\label{eq:fit}
\end{equation}
with $a = 0.0108805$, $b=-1.0133\times10^{-4}$, $c= 0.02228$, and $d=1.84558\times10^{-4}$ and
where $\zeta=eB/\Lambda_{QCD}^2$. We have used $\Lambda_{QCD}=300$ MeV.

\begin{figure}[t]
\centering
    \includegraphics[width=0.9\linewidth,angle=0]{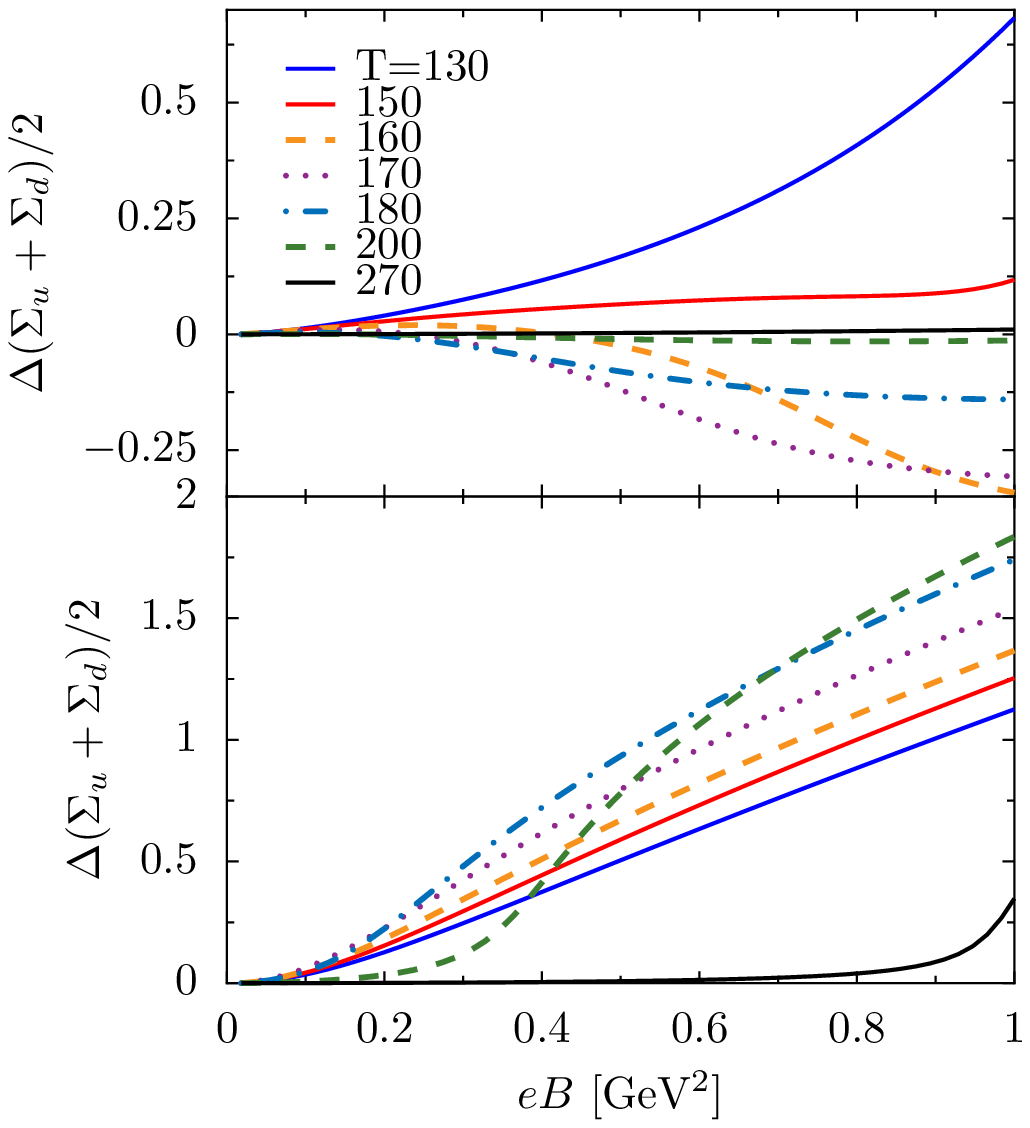}
\caption{
The light chiral condensate $\Delta(\Sigma_u+\Sigma_d)/2$ as a function of 
$eB$, for several values of temperature in MeV, in the NJL model, with
a magnetic field dependent coupling $G_s(eB)$ from Eq. (\ref{eq:fit})
(top) and a constant coupling $G_s^0$ (bottom).
}
\label{fig:chiral_cond_soma}
\end{figure}
In \cite{Miransky:2002rp}, the authors have shown that in the presence of an
external magnetic field and in the intermediate regime, corresponding to  
an energy scale below the magnetic field scale but larger than the dynamical 
quark mass,  the gluon acquires a mass $M_g^2\propto\alpha_s|eB|$.
Thus in this  limit of interest precisely at the chiral symmetry
restoration transition, we have $G_s\propto\alpha_s/M_g^2\propto1/eB$.
The  above polynomial form  insures that for $eB\rightarrow\infty$, $G_s$ goes as $1/eB$.

\begin{figure}[t]
\centering
    \includegraphics[width=0.85\linewidth,angle=0]{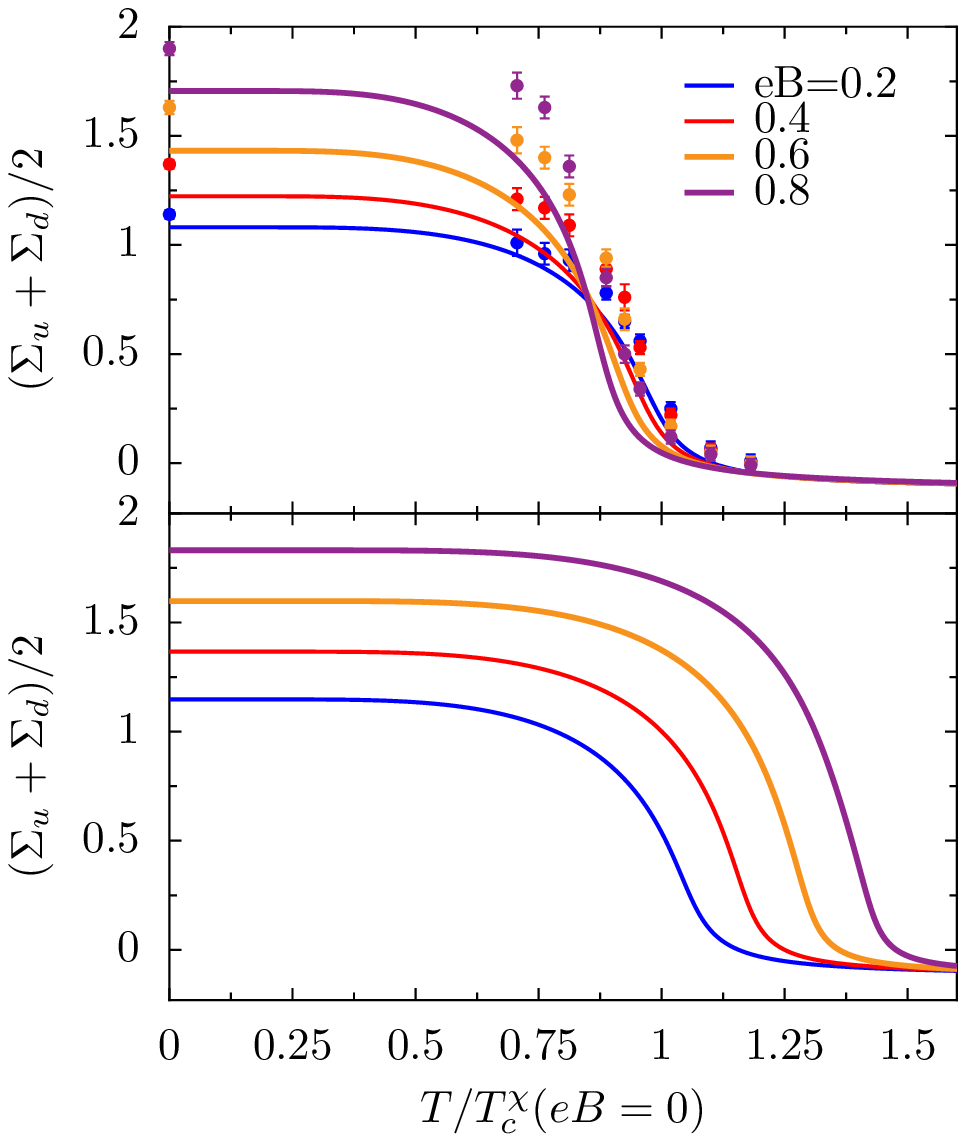}
\caption{
The light chiral condensate $(\Sigma_u+\Sigma_d)/2$ as a function of 
temperature, for several values of $eB$ in GeV$^2$, in the NJL model, with a 
magnetic field dependent coupling $G_s(eB)$ from Eq. (\ref{eq:fit})
compared with LQCD results \cite{bali2012PRD} (top), and a constant coupling $G_s^0$ (bottom).
The LQCD data was renormalized by $T_c^\chi(eB=0)=160$ MeV \cite{bali2012PRD} and the NJL model
results by $T_c^\chi(eB=0)=178$ MeV.
}
\label{Fig:cond_soma_T}
\end{figure}
\begin{figure}[t!]
\centering
    \includegraphics[width=0.85\linewidth,angle=0]{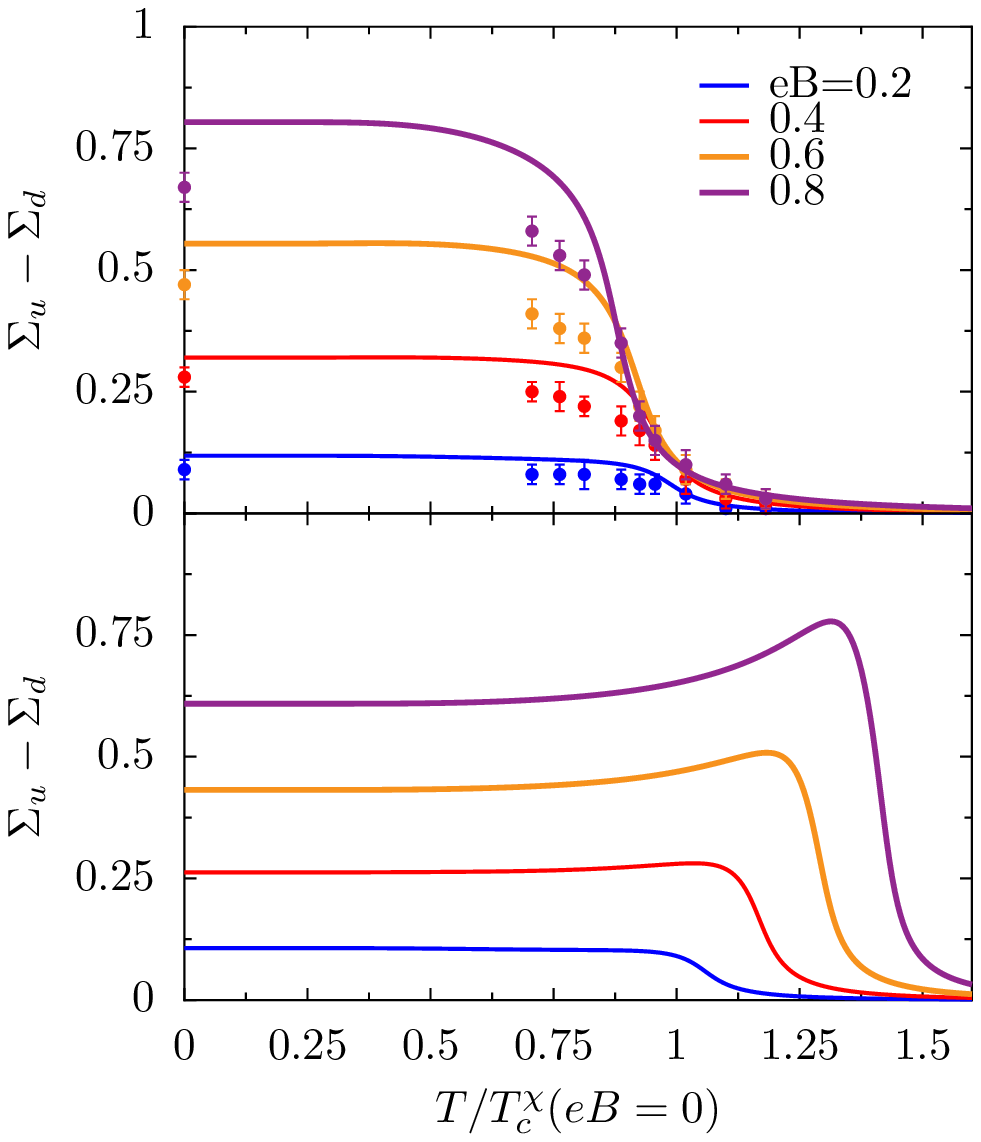}
\caption{
The chiral condensate difference $\Sigma_u-\Sigma_d$ as a function of 
temperature, for several values of $eB$ in GeV$^2$, in the NJL model, 
calculated with a magnetic field dependent coupling $G_s(eB)$ [Eq. (\ref{eq:fit})]
compared with LQCD results \cite{bali2012PRD} (top), and a constant coupling $G_s^0$ (bottom).
The LQCD data was renormalized by $T_c^\chi(eB=0)=160$ MeV \cite{bali2012PRD} and the NJL model
results by $T_c^\chi(eB=0)=178$ MeV.
}
\label{Fig:cond_soma_dif_T}
\end{figure}
\begin{figure}[t!]
\centering
    \includegraphics[width=0.85\linewidth,angle=0]{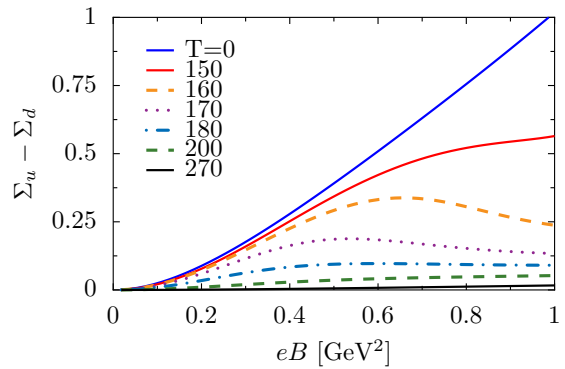}
\caption{
The chiral condensate difference $\Sigma_u-\Sigma_d$ as a function of 
$eB$, for several values of temperature in MeV, in the NJL model, 
with a magnetic field dependent coupling $G_s(eB)$, Eq. (\ref{eq:fit}).
}
\label{fig:chiral_cond_dif}
\end{figure}

In the following, we focus on the order parameter for the chiral transition, and, according to \cite{bali2012PRD},
we define the change of the light quark condensate due to the magnetic field as
\begin{equation}
\Delta\Sigma_{f}(B,T)=\Sigma_{f}(B,T)-\Sigma_{f}(0,T),
\label{Delta}
\end{equation}
with
\begin{equation}
\Sigma_{f}(B,T)=\frac{2M_f}{m_\pi^2f_\pi^2}
\left[\left\langle \bar{q}_{f}q_{f}\right\rangle(B,T) \right.
- \left. \left\langle \bar{q}_{f}q_{f}\right\rangle(0,0) \right]+1
\end{equation}
where the factor $m_\pi^2f_\pi^2$ in the denominator contains the pion
mass in the vacuum ($m_\pi=135$ MeV) and (the chiral limit of the) pion
decay constant ($f_\pi=87.9$ MeV) in NJL model. The behavior of the quark 
condensates with the magnetic field is shown in Figs. 
\ref{fig:chiral_cond_soma}-\ref{fig:chiral_cond_dif}.

In Fig. \ref{fig:chiral_cond_soma}, the change of the renormalized condensate 
$\overline{\Delta \Sigma}=\Delta(\Sigma_{u}+\Sigma_{d})/2$ as a function
of the magnetic field intensity at several temperatures is shown for 
$G_s(eB)$ defined in Eq. (\ref{eq:fit}) (top panel) and $G_s^0$ (bottom panel).
The average of the light condensates calculated with $G_s(eB)$  shows
the same behavior as LQCD calculations:
 at low and high temperatures the magnetic field enhances 
the condensates but at temperatures near the pseudocritical chiral transition temperature, the
magnetic field suppresses the condensates. Using $G_s=G_s^0$, an enhancement is predicted
at any temperature \cite{Ferreira:2013tba}: the magnetic catalysis is
the result of an enhancement of the spectral density at low energies
which increases the number of participants in the chiral  condensate.

If $G_s=G_s^0$, and for $T<T_c^{\chi}(eB=0)$, $\overline{\Delta \Sigma}$ 
increases with $eB$ due to the magnetic catalysis effect, 
being its value even larger as the temperature is higher \cite{Ferreira:2013tba}. 
On the other hand, when $T>T_c^{\chi}(eB=0)$ we are in the region where the partial 
restoration of chiral symmetry already took place. In this region there are two 
competitive effects: the partial restoration of chiral symmetry,  that  prevails at 
lower values of $eB$, making the condensate average approximately zero,  and the 
magnetic catalysis, that becomes dominant as the magnetic field increases.
When the strength of the interaction decreases as $eB$ increases, the coupling of a 
quark-antiquark pair interaction is weakened leading to the occurrence of an earlier 
partial restoration of chiral symmetry, so this effect is dominant preventing 
the magnetic catalysis to occur.

These same conclusions are obtained from  Fig. \ref{Fig:cond_soma_T}
where the  average light quark condensate is plotted as functions of $T$ for several
values of $eB$. The lattice results extracted from \cite{bali2012PRD} have also been 
included in the top panel together with the results obtained with  $G_s=G_s(eB)$ from 
Eq. (\ref{eq:fit}). The qualitative agreement between both calculations is quite good, 
with the main features being reproduced by the NJL model. A very different behavior is
obtained with a constant coupling $G_s^0$, see Fig. \ref{Fig:cond_soma_T} bottom,  
where the transition for larger values of $eB$  occurs for larger temperatures.

In Figs. \ref{Fig:cond_soma_dif_T}  and  \ref{fig:chiral_cond_dif} the difference between 
light quark condensate are plotted, respectively,  as a function of $T$ for several
values of $eB$, and as a function of $eB$ for several temperatures. The lattice results  
from \cite{bali2012PRD} have been included in Fig. \ref{Fig:cond_soma_dif_T} (top) together 
with the results for $G_s=G_s(eB)$. For comparison we also show the  results for $G_s=G_s^0$
(bottom).

The bumps  present in curves obtained with $G_s=G_s^0$  around the transition
temperatures, a characteristic of the NJL with constant coupling,
do not appear when $G_s(eB)$ is used, and a reasonable agreement with
the LQCD results is achieved.
As pointed out in \cite{Ferreira:2013tba}, these bumps are the result of a stronger 
magnetic catalysis effect for the $u$ quark, due to its larger electric charge, 
(the larger the magnetic field the larger the difference between $u$ and $d$ condensates, and 
the respective chiral transition temperatures), 
being this feature particularly strong close to the transition temperature, where the curves 
for stronger fields have a larger bump.
When $G_s=G_s(eB)$, the effect due to the partial restoration of chiral symmetry will prevail
over the magnetic catalysis due to a weaker interaction and, as already pointed out, the 
bumps will disappear in accordance with lattice results.
In Fig. \ref{fig:chiral_cond_dif} the condensate difference $\Sigma_u-\Sigma_d$ is plotted as 
a function of $eB$ for several temperatures and it is clearly seen that it always decreases with 
the temperature.

We next analyze the $T-eB$ phase diagram obtained within the NJL with
a coupling dependent on the magnetic field.
The parametrization $G_s(eB)$ was obtained using the available lattice results for
the pseudocritical temperatures  in the range $0<eB<1$ GeV$^2$. 
The calculated pseudocritical temperatures  are shown in Fig. \ref{Fig:T_c} for a range of 
magnetic field intensities larger than the one used in the fit.
For $eB\approx1.1$ GeV$^2$, the pseudocritical temperature starts to increase with $eB$. 
This behavior was also obtained in lattice calculations \cite{Ilgenfritz:2013ara},
which also predict magnetic catalysis at high values of $eB$.
For $eB\approx1.25$ GeV$^2$  a chiral first order phase transition appears.
The LQCD as well as the NJL results from Fig. \ref{Fig:cond_soma_T} show that the average
chiral condensate slope increases with increasing magnetic field. Thus, if this
behavior persists for high magnetic fields, it is expected also from the LQCD results, 
that at some critical $eB$, the transition turns into a first order.

\begin{figure}[t]
\centering
    \includegraphics[width=0.85\linewidth,angle=0]{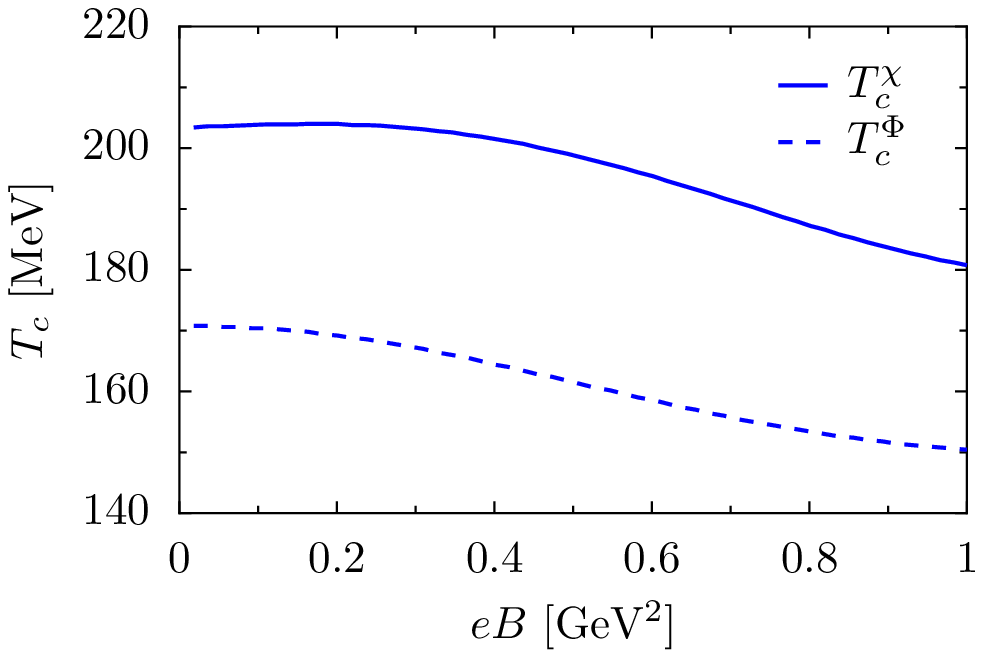}
\caption{
The chiral and deconfinement transitions temperatures as a function of $eB$
in the PNJL, using the
magnetic field dependent coupling $G_s(eB)$ [Eq. (\ref{eq:fit})].
}
\label{fig:pnjl_tc}
\end{figure}

\subsection{PNJL model}

In the present section, we consider the PNJL model. In this model the
the quark degrees of freedom are coupled to a Polyakov loop field
which allows us to simulate a  confinement-deconfinement
phase transition at finite temperature. 
Several studies about the deconfinement and chiral symmetry restoration
of hot QCD matter in an external magnetic field have recently been made
\cite{Ferreira:2013oda,Costa:2013zca,Fu:2013ica}.
Now, we will take for the scalar coupling parameter the same magnetic field 
dependent parametrization obtained in the previous section [Eq. (\ref{eq:fit})]. 

Next we discuss the effect of the magnetic field on the Polyakov loop and on the quark
condensates within this model.
\begin{figure}[t!]
\centering
    \includegraphics[width=0.85\linewidth,angle=0]{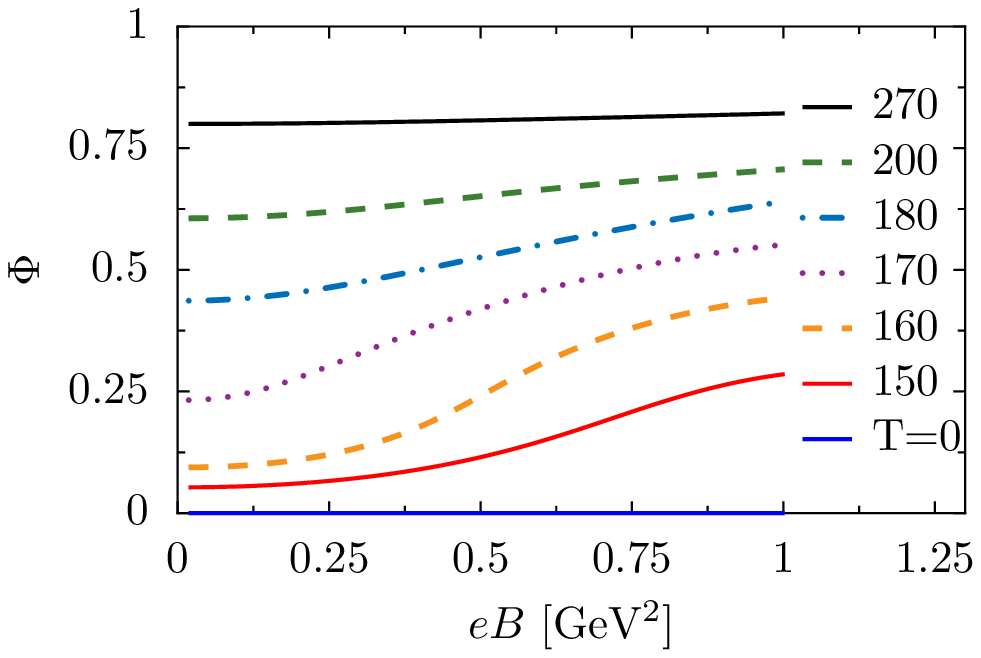}\\
    \includegraphics[width=0.85\linewidth,angle=0]{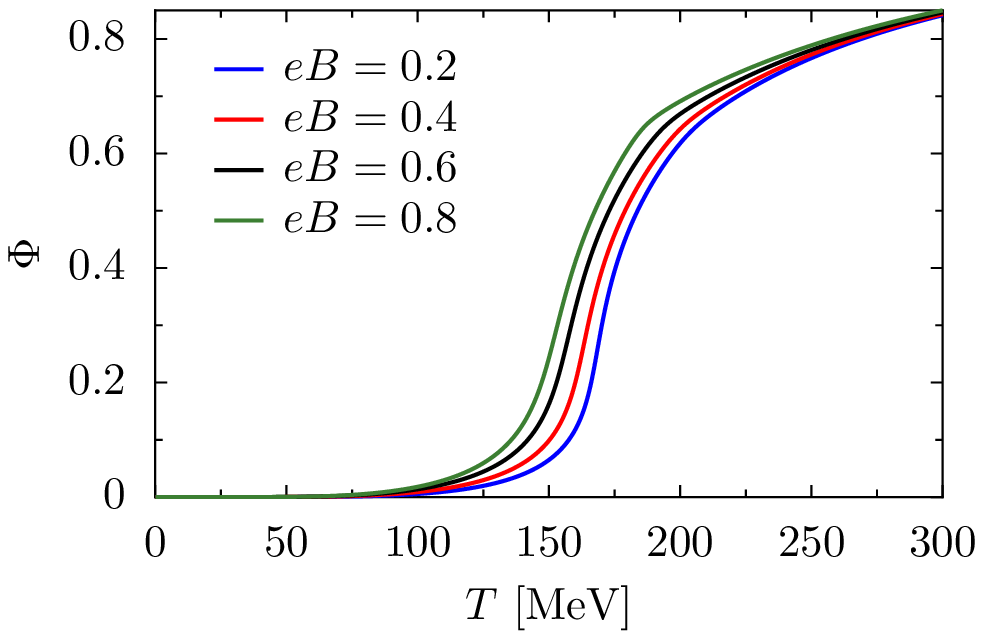}
\caption{
The value of the Polyakov loop versus $eB$ for several values of $T$ (MeV) 
(top) and versus $T$ for several values of $eB$ in GeV$^2$ (bottom).
}
\label{fig:polyakov}
\end{figure}

Some remarks are in order concerning the applicability of the PNJL model. It should be noticed that
in this model, besides the chiral point-like coupling between quarks, the gluon 
dynamics is reduced to a simple static background field representing the Polyakov loop.
As referred in Sec. \ref{sec:model}, we consider the parameter $T_0$ in the  
Polyakov loop $T_0=210$ MeV, which takes into account the quark backreaction 
and reproduces the lattice deconfinement pseudocritical temperature $170$ MeV. 
As shown in the following, we obtain within the PNJL model several features 
discussed in the previous section, e. g.  the deconfinement transition and 
the chiral transition pseudocritical temperatures are both  decreasing functions with $eB$  
until a limiting magnetic field of $\sim 1$ GeV$^2$, as in LQCD, see 
Fig. \ref{fig:pnjl_tc}. 
Due to the existing coupling between the Polyakov loop field and  quarks within PNJL, 
the $G_s(eB)$ does not only affect the chiral transition but also the deconfinement 
transition. 
\begin{figure}[t!]
\centering
    \includegraphics[width=0.85\linewidth,angle=0]{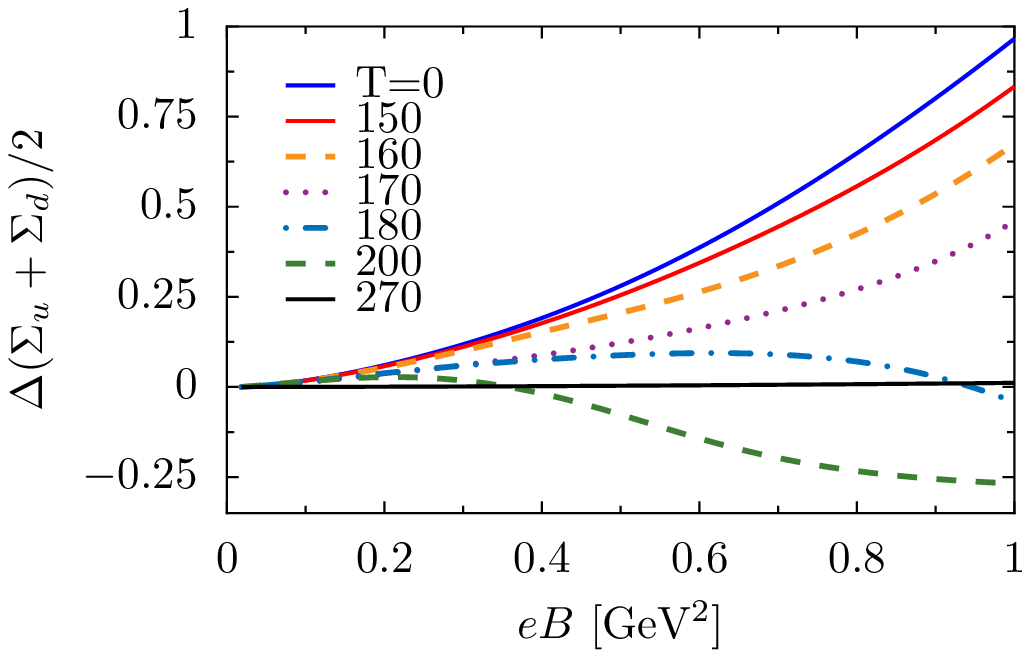}
\caption{
The light chiral condensate $\Delta(\Sigma_u+\Sigma_d)/2$ as a function of 
$eB$, for several values of $T$ in MeV, in the PNJL model.
}
\label{fig:pnjl_sum}
\end{figure}

The effect of the magnetic field on the Polyakov loop is more clearly
seen in Fig. \ref{fig:polyakov} where $\Phi$ is plotted as a function
of the magnetic field intensity for different values of the
temperature (top), and as a function of temperature, for several
magnetic field strengths (bottom). The suppression of the condensates
achieved by the  magnetic field dependence of the coupling parameter
translates into  an increase of the Polyakov loop. The effect of the
magnetic field on $\Phi$ is stronger precisely for the temperatures close
to the transition temperature, see Fig. \ref{fig:polyakov} (top), in
close agreement with the LQCD results \cite{endrodi2013}.

\begin{figure}[t!]
\centering
    \includegraphics[width=0.85\linewidth,angle=0]{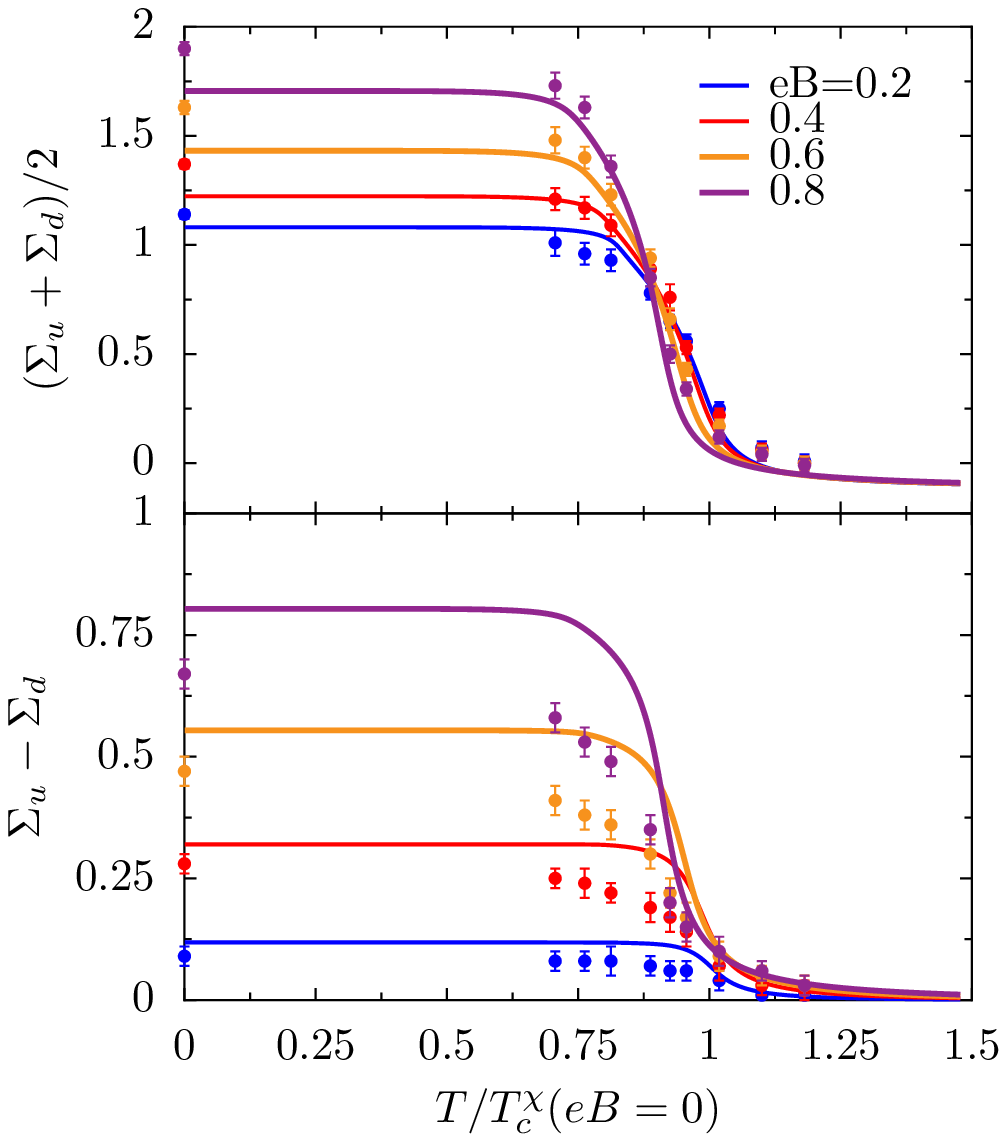}
    \caption{
The average $(\Sigma_u+\Sigma_d)/2$ (top) and the difference $(\Sigma_u-\Sigma_d)$ (bottom)
of the light chiral condensates as a function of 
temperature, for several values of $eB$ in GeV$^2$, and the LQCD results \cite{bali2012PRD}.
The LQCD data was renormalized by $T_c^\chi(eB=0)=160$ MeV \cite{bali2012PRD} and the PNJL model
results by $T_c^\chi(eB=0)=203$ MeV.
}
\label{fig:pnjl_cond2}
\end{figure}
In Fig. \ref{fig:pnjl_sum} we plot the average chiral condensate 
$\Delta(\Sigma_u+\Sigma_d)/2$ 
as a function of $eB$, for several temperatures. As in the LQCD \cite{bali2012PRD}, 
for temperatures smaller and higher than the transition temperatures, the model predicts 
a monotonously increase with $eB$, and for temperatures near the transition temperature,
a nonmonotonic behavior is obtained. Thus, for $T\approx T^\Phi_c$, at magnetic
field intensities higher than some pseudocritical value, the condensates are suppressed 
by the presence of the magnetic field.

In Fig. \ref{fig:pnjl_cond2} the chiral condensate sum  $(\Sigma_u+\Sigma_d)/2$ and  
the chiral condensate difference $\Sigma_u-\Sigma_d$  are plotted as a function of 
the  temperature,  renormalized by the pseudocritical temperature at zero magnetic 
field, for several magnetic field strengths and compared with the LQCD results
from \cite{bali2012PRD}.  Just as already obtained for NJL, general
features of the LQCD results are reproduced.

We observe that SU(3) symmetry of the pointlike effective interactions between quarks 
is assumed in the magnetic background, however the comparison with the LQCD results for 
the difference in the quark condensates in Fig. \ref{fig:pnjl_cond2} bottom, suggests 
that the up quark interaction is depleted with respect to the down quark one. 
That, seems reasonable as the effect of the magnetic field on the up quark is larger than 
in the down quark, and therefore the interaction between the up quarks should decrease 
with respect to the down quarks as the magnetic field increases. A more detailed calculation 
should account for magnetic SU(3) flavor breaking and it is postponed for a future work.

\subsection{$G_s(eB)$ versus $T_0(eB)$}

\begin{figure}[t!]
\centering
    \includegraphics[width=0.85\linewidth,angle=0]{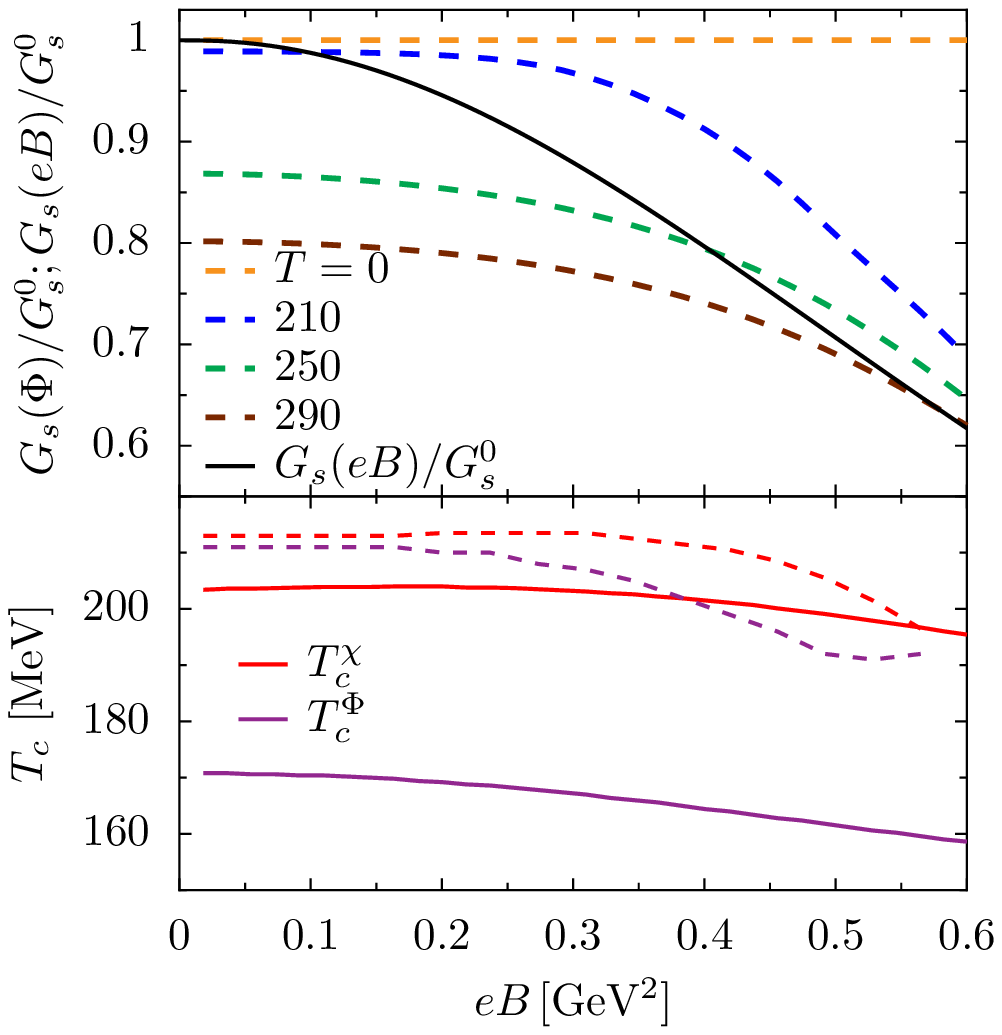}\\
    \includegraphics[width=0.85\linewidth,angle=0]{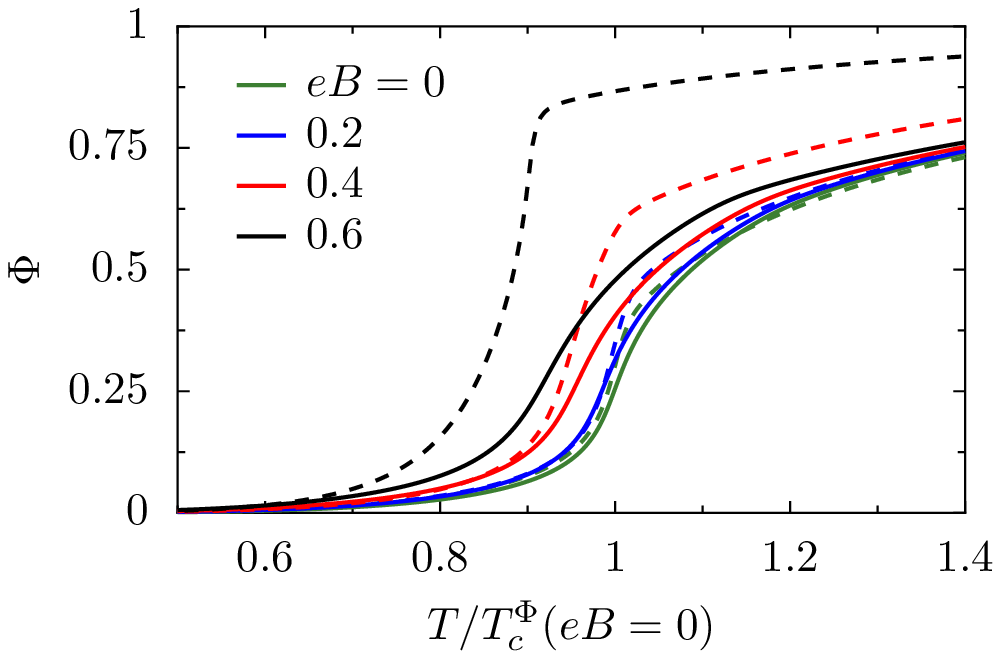}
    \caption{Comparison between present work (full lines) and
      as results obtained within  EPNJL with $T_0(eB)$
      \cite{Ferreira:2013tba}. Top: The scalar coupling $G_s$ versus
      the magnetic field, the black full line is the parametrization
      defined in (\ref{eq:fit}) and plotted in Fig. \ref{fig:fit}; middle: the chiral and deconfinement
      pseudocritical temperatures versus the magnetic field; bottom:
      the Polyakov loop versus the temperature renormalized by the
      deconfinement pseudocritical temperature $T_c^\Phi$ for  $eB=0$,
      respectively,  171 MeV  [PNJL with $G_s(eB)$] and 214 MeV [EPNJL with $T_0(eB)$].
}
\label{epnjl}
\end{figure}

In Ref. \cite{Ferreira:2013tba} the possibility of describing the IMC 
effect within the models PNJL and entangled PNJL (EPNJL)  by including 
a  magnetic field dependent parameter $T_0(eB)$ in the parametrization 
of the Polyakov potential was studied. The main argument in
favor was that backreaction effects on the Polyakov loop due to the
presence of a strong magnetic field should introduce screening effects
leading to a reduction of the pseudocritical transition
temperatures. A magnetic field dependent  effective Polyakov potential
could indeed describe the IMC effect but only within EPNJL. Neither
the PNJL model
\cite{Ferreira:2013tba}  nor  the  two-flavor thermal quark-meson model 
\cite{Fraga:2013ova} were able to obtain the IMC effect with  a $T_0$ 
parameter dependent on the magnetic field. These results are in accordance
with the results of the present work: in the EPNJL the coupling $G_s$
depends on the Polyakov loop, and, therefore, at the crossover when
the Polyakov loop increases the coupling $G_s$ becomes weaker. This is
shown in Fig. \ref{epnjl} top panel, where the coupling $G_s[\Phi(T)]$ of Ref.
\cite{Ferreira:2013tba} is plotted for
several temperatures (dashed curves)  and, for comparison, the parametrization $G_s(eB)$
given in Eq. (\ref{eq:fit})  and Fig. \ref{fig:fit} is also
included (full black line). 
It is interesting to realize that in the range  $eB<0.6$ GeV$^2$, 
the curve obtained for $T=210$ MeV, which is close to the
deconfinement  pseudocritical  temperature ($T_c^\Phi=214$ MeV), has a
behavior in accordance  with the results of the present work. 
Within the PNJL no IMC effect was obtained because the parameter $T_0(eB)$ 
does not affect the coupling $G_s$.

In Fig. \ref{epnjl} middle and bottom panels, we compare the results obtained
in the present work with the ones of \cite{Ferreira:2013tba} for the
pseudocritical  temperatures and the Polyakov loop. 
In \cite{Ferreira:2013tba}, the pseudocritical temperatures have a much 
flatter behavior at small values of the magnetic field reflecting the 
softer decrease of the coupling $G_s$ for small values of the magnetic 
field shown in Fig. \ref{epnjl} top panel.  Also, within the EPNJL
with  $T_0(eB)$ the difference between the pseudocritical temperatures
$T_c^\chi$ and $T_c^\Phi$  is much smaller because the Polyakov loop and
the quark condensates are strongly coupled. For $eB$=0 these
temperatures are almost coincident, but a finite strong magnetic field
destroys this coincidence. PNJL does not have this feature and a
$G_s(eB)$ coupling is not changing its normal behavior predicting
different temperatures for  $T_c^\chi$ and $T_c^\Phi$.
On the other hand, the effect of the
parametrization $T_0(eB)$ on the Polyakov loop  within EPNJL is much stronger than
the one obtained in the present work, which is an indirect effect
occurring due to the dependence of the quark distributions, 
Eq. (\ref{eq:z}), on the Polyakov loop.

\section{Conclusions}

In the present work we study  quark condensates and  the QCD phase diagram at 
zero chemical potential and finite temperature subject to an external magnetic 
field within the NJL and PNJL models. 

We have shown that recent results from LQCD, for quark matter
in the presence of an external magnetic field, can be reproduced
within NJL/PNJL models, if a magnetic field dependent coupling constant 
is used. A decreasing magnetic field dependent four quark coupling is essential,
within effective quark models, to mimic the expected running of the
coupling constant with the magnetic field strength and to incorporate
the backreaction of the sea quarks in order to explain the IMC.

We have calculated the $G_s(eB)$ coupling constant in the NJL model, 
that reproduces the qualitative behavior of chiral pseudocritical temperature given by LQCD. 
All the qualitative results predicted by LQCD, can be reproduced using 
the calculated $G_s(eB)$ coupling: 
(a) the nonmonotonic behavior of the average condensate $\Delta(\Sigma_u+\Sigma_d)/2$ 
as a function of $eB$ is obtained  in the temperature region of the chiral transition; 
(b) the Polyakov loop increases with  $eB$, and this increase is stronger
for temperatures in the temperature region of the chiral transition;
(c) the difference between the $u$ and $d$ quarks decreases
monotonically with  temperature, contrary to the
prediction of the NJL and PNJL with constant couplings that predict an
increase of this difference until the transition temperature. Furthermore, 
LQCD results suggest that the SU(3) symmetry of the pointlike effective 
interaction between quarks should be broken in the magnetic environment.

\vspace{0.25cm}
{\bf ACKNOWLEDGMENTS}:
This work was partially supported by Project No. PTDC/FIS/ 113292/2009
developed under the initiative QREN financed by the UE/FEDER through the
program COMPETE: "`Programa Operacional Factores de
Competitividade"', and by Grant No. SFRH/BD/51717/2011.
We are also grateful to J. Moreira for useful discussions. 
T.F. thanks  the Conselho Nacional de Desenvolvimento Cient\'ifico e Tecnol\'ogico (CNPq) 
and  T.F. and O.L. thank the Funda\c c\~ao de Amparo a Pesquisa do Estado de S\~ao Paulo (FAPESP).

\end{document}